\documentclass[acus]{JAC2003}

\usepackage{graphicx}
\usepackage{booktabs}
\usepackage{amssymb}
\usepackage{wrapfig}

\begin{document}
\title{THE FIRST 1 1/2 YEARS OF TOTEM ROMAN POT OPERATION AT LHC}

\author{M. Deile, G. Antchev, I. Atanassov, V. Avati, J. Baechler, K. Eggert, 
J. Ka\v{s}par, F. Lucas Rodriguez,\\ J. Morant, H. Niewiadomski, 
E. Radermacher, F. Ravotti, G. Ruggiero, H. Sabba, W. Snoeys\\
on behalf of the TOTEM Collaboration;\\
R.B. Appleby, R. Assmann, R. Bruce, M. Dupont, M. Dutour, B. Farnham, 
X. Pons, S. Ravat,\\ 
S. Redaelli, M. Sapinski, G. Valentino, D. Wollmann}

\maketitle

\begin{abstract}
Since the LHC running season 2010, the TOTEM Roman Pots (RPs) are fully operational and serve for collecting elastic and diffractive proton-proton scattering data. Like for other moveable devices approaching the high intensity LHC beams, a reliable and precise control of the RP position is critical to machine protection. After a review of the RP movement control and position interlock system, the crucial task of alignment will be discussed.
\end{abstract}

\section{INTRODUCTION}
The TOTEM Roman Pot system~\cite{TOTEM} 
has been designed for measuring very forward protons originating from elastic 
or diffractive scattering processes. With different beam optics, four-momentum
transfers down to $\sim 10^{-3}\,\rm GeV^{2}$ can be reached,
corresponding to scattering angles of a few $\mu$rad. The proton detection is
accomplished by stacks of 10 silicon sensors placed in movable beam-pipe 
insertions -- so-called ``Roman Pots'' (RPs) -- that can approach the 
beam. In some special runs the distance from the beam centre has been smaller than 1\,mm. The system consists of 24 RPs
arranged in 4 stations at 147\,m and 220\,m on either side of LHC's IP5.
Each
station is composed of 2 units separated by a distance of 2 -- 5\,m. A unit 
consists of 3 RPs, two approaching the outgoing beam vertically and one 
horizontally, allowing for a partial overlap between horizontal and vertical
detectors, which is useful for the alignment.

%

\section{THE ROMAN POT MOVEMENT CONTROL SYSTEM}

\subsection{System Architecture}
The RP movement control system~\cite{usecases} 
has been derived from the one of the LHC collimators.
The top level user interface, 
a modified version of the Central Collimation Application~\cite{cca}, 
is operated from the CCC. It allows the operator to send RP position requests 
and critical position limits to the middle 
level control (see below) and to read back the measured positions. A graphical 
display shows the time evolution of requested and measured RP positions, 
the position limits, and the dose rates measured by Beam Loss Monitors 
downstream of the RPs (see example screenshot in Fig.~\ref{fig:interlocktest}).
The middle level is the FESA-based~\cite{fesa} Collimator Supervisor System~\cite{middle}, 
providing the 
link between the Collimation Application and the low level motor control running
on a PXI~\cite{pxi} crate. The PXI CPU controls the stepping motors of the 
individual RPs and manages the position read-outs from the stepping motor 
encoder (step counter) and from the independent LVDT system (see next subsection).
The PXI computer also interfaces to FPGA cards which independently from the CPU
accomplish the 
interlock-relevant comparison between actual RP positions and their critical 
limits (see later subsection). The FPGA cards have direct connections from the 
microswitches and the LVDTs and to the beam interlock logic card.

Note that from the TOTEM control room no RP movements can be triggered,
except emergency retractions via mechanical springs. 
This policy enforces centralisation of controls in the CCC 
and enhances communication between TOTEM and the LHC operators.
However, the TOTEM Detector Control System monitors passively all operations.

\subsection{Microswitches and Position Measurements}
\begin{wrapfigure}{r}{50mm}
    \vspace*{-3mm}
    \centering
    \hspace*{-6mm}
    \includegraphics[width=55mm]{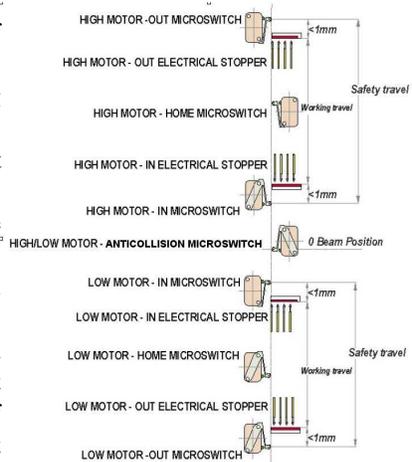}
    \caption{Schematic overview of the RP microswitches tagging some particular 
pot positions and defining the travel range.
}
    \label{fig:microswitches}
    \vspace*{-3mm}
\end{wrapfigure}
The RP mechanics has a series of microswitches~\cite{review} 
(Fig.~\ref{fig:microswitches}) 
tagging certain positions in the RPs' movement range and triggering 
appropriate responses. The IN and OUT electrical stoppers define the range of 
motion and stop the pot at its first point of electrical contact with their 
spring test probes. These probes have an elastic travel of 3.18 mm beyond the 
first point of contact. The OUT microswitch beyond the electrical stopper acts
as a redundant motor stopper.
The OUT stopper also serves as absolute position reference; its 
on-off transition point has been surveyed by laser calibration. 
The HOME switch is relevant for interlock purposes. The HOME flag is active 
in all retracted pot positions beyond the first contact point with the HOME 
switch. The ANTICOLLISION switch stops the motors if a top and a bottom pot
come too close to each other.

The RP system has two redundant position measurements: 1. The motor encoders, 
i.e. step counters, are used for active movement control. Each step of the 
stepping motor corresponds to about 
5\,$\mu$m. Due to a small position dependence of the step size, 
step calibration curves are measured for each RP at least once per year.
The encoder measures the position relative to the OUT electrical stopper; it
has no absolute position memory.
In a fast extraction of a pot with the springs, the mechanical coupling between
the pot and its motor is removed, leading to a loss of the absolute position 
information. Hence, to ensure a correct calibration, each RP insertion is 
preceded by an encoder reset at the OUT stopper.
2. The LVDTs~\cite{lvdt} provide an absolute position measurement but are 
subject to small drifts and need to be periodically recalibrated. These 
measurements are used for the position interlock and as an independent 
cross-check of the encoders.

\vspace*{-1.5mm}
\subsection{TOTEM Beam Interlock Logic}
Like all LHC experiments and many machine components, the TOTEM RP system 
has an input to the LHC beam interlock system to avoid beam operation when a
RP is in a position where it would represent a danger. 

This logic is implemented in the programmable TOTEM interlock card that
receives beam mode information (SMP parameters) from the General Machine Timing
(GMT), RP position validity information from the FPGA cards 
comparing each RP's LVDT position with its beam-mode dependent critical 
limits, and 
simple electrical contact information from the RP Home microswitches.
The output signals, INJECTION\_PERMIT and USER\_PERMIT are transmitted by 
hardware links to the beam user interfaces (``CIBF'' and ``CIBU''); copies are
sent to the Detector Control System. The key functions of the interlock 
logic~\cite{usecases,interlock} are the following: 1. Withdrawal of INJECTION\_PERMIT:
Prevention of any beam injection
when at least one RP does not touch its HOME microswitch. 
2. Withdrawal of USER\_PERMIT: Beam dump and automatic RP extraction  
if either (a) at least one 
RP is away from its HOME switch in a beam mode where RP operation is not 
allowed, or (b) at least one RP is outside the position range defined by 
beam-mode dependent critical limits.

\begin{figure}[h!]
    \vspace*{2mm}
    \centering
    \hspace*{-1mm}
    \includegraphics[width=83mm]{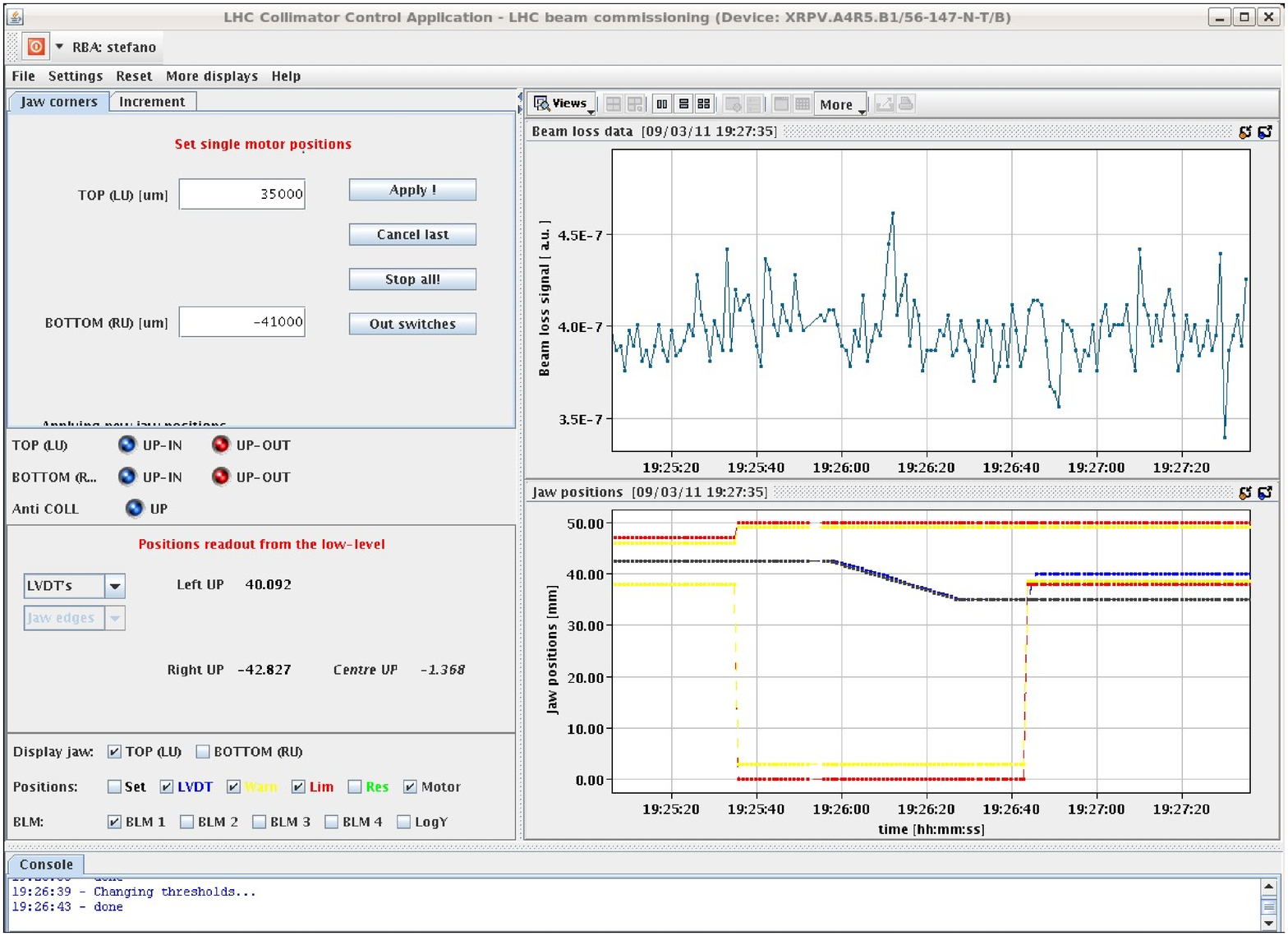}
    \vspace*{-4mm}
    \caption{RP movement sequence during the interlock tests 2011:
black = motor step counter position, blue = LVDT position, 
red = outer and inner dump limits, yellow = outer and inner warning limits. 
At 19:26:40 the inner limits are changed such that  the 
RP position becomes illegal. Consequently the pot is automatically retracted. 
The LVDT correctly indicates the new position ($\sim$39.7\,mm). The step 
counter, however, stays at 37\,mm because in the emergency 
extraction with the springs the motor coupling is lost.
A reset at the mechanical reference point (OUT stopper) is needed.
}
    \label{fig:interlocktest}
    \vspace*{-4mm}
\end{figure}

An example sequence of an interlock test performed with the Collimation 
Application in the CCC is shown in Fig.~\ref{fig:interlocktest}.

\vspace*{-4.5mm}
\section{ROMAN POT ALIGNMENT}
\vspace*{-1mm}

\subsection{Beam-Based Alignment}
\begin{figure}[h!]
    \vspace*{-4mm}
    \centering
    \includegraphics[width=75mm]{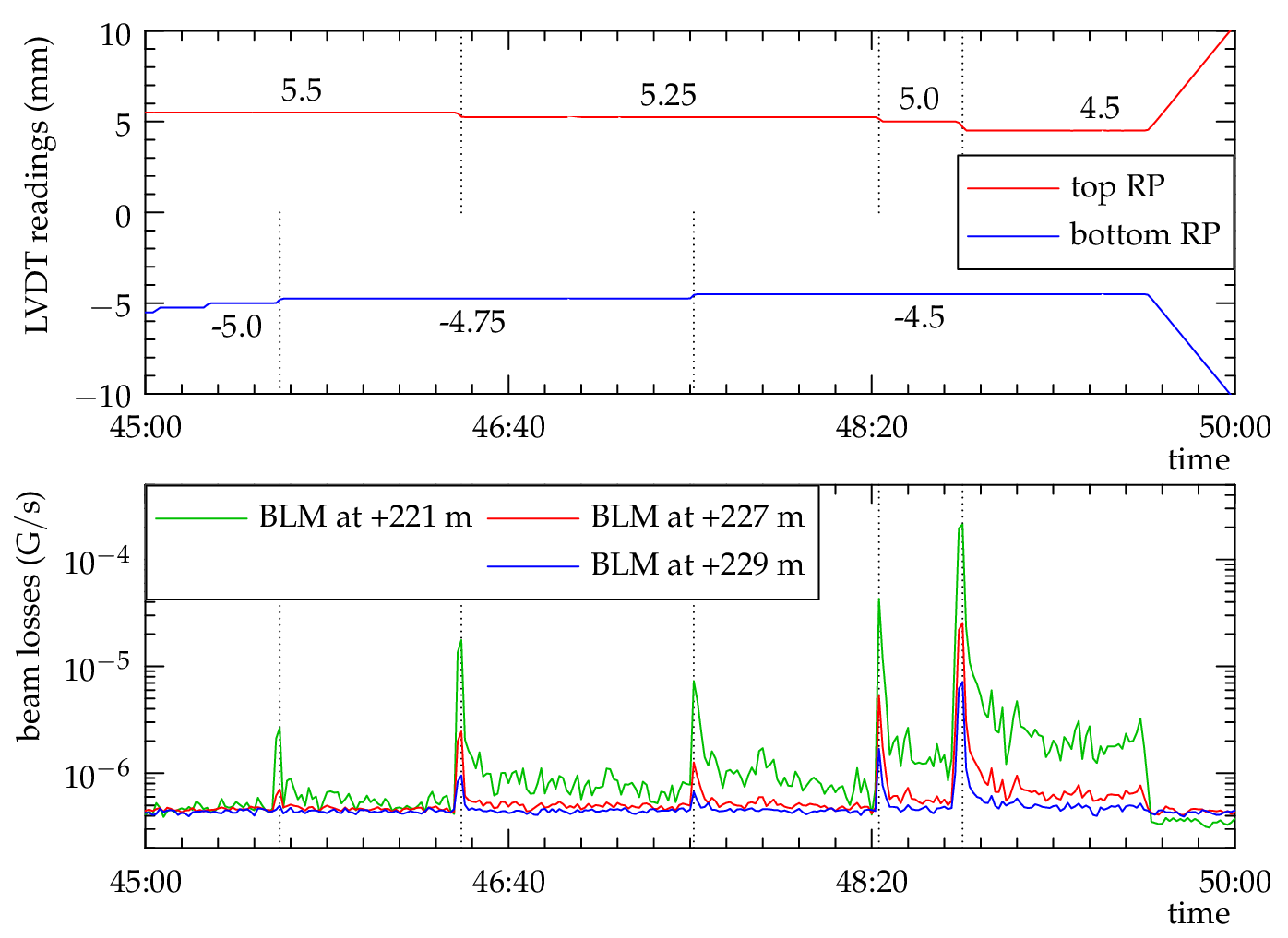}
    \vspace*{-5mm}
    \caption{Beam-based alignment sequence for a vertical RP pair in 2010.
In this early alignment exercise the step size was large (250\,$\mu$m). 
Nowadays, 10\,$\mu$m are used for better precision. The integrated 
dose in the BLM spikes
was also used to calibrate the BLM response~\cite{blm}.}
    \label{fig:collim}
\end{figure}

The beam-based alignment of the RPs is performed with a 
technique established for the collimators~\cite{collim}: a reference 
beam edge is produced by closing the horizontal and vertical primary 
collimators (TCPs) until they scrape the beam halo at an aperture 
$n_{\rm TCP} \sigma_{\rm TCP}$, where $\sigma_{\rm TCP}$ is the beam size at the 
collimator, calculated for nominal optics functions and for a given normalised 
emittance. This beam edge is then used as a reference for 
the RPs: the pot approaches the circulating beam in steps as small as 
10\,$\mu$m. When it reaches the edge, a beam loss is measured in a Beam Loss 
Monitor (BLM) downstream (Fig.~\ref{fig:collim}). 
In this condition, the RP and the primary 
collimator are at the same normalised distance $n_{\rm TCP}$ 
from the beam orbit with an accuracy given by the step size. 
At the contact position, the RP scrapes slightly into the beam.
To refine the scraped beam edge, the primary collimator in the 
appropriate plane is moved further towards the beam until it becomes 
again the aperture bottleneck. 
The average of the two normalised reference distances $n_{\rm TCP}$ of 
the primary collimators is used to compute the normalised distance 
$n_{\rm RP}$ of the RP. 

Note that this procedure only establishes the distance between the \textit{thin window} 
of the RP and the beam centre. 
The alignment of the different \textit{silicon detector planes}, both relative 
to each other and to the beam, is 
determined by software, using real particle data.

\subsection{Software Alignment}
\begin{wrapfigure}{r}{0pt}
    \centering
    \includegraphics[width=40mm]{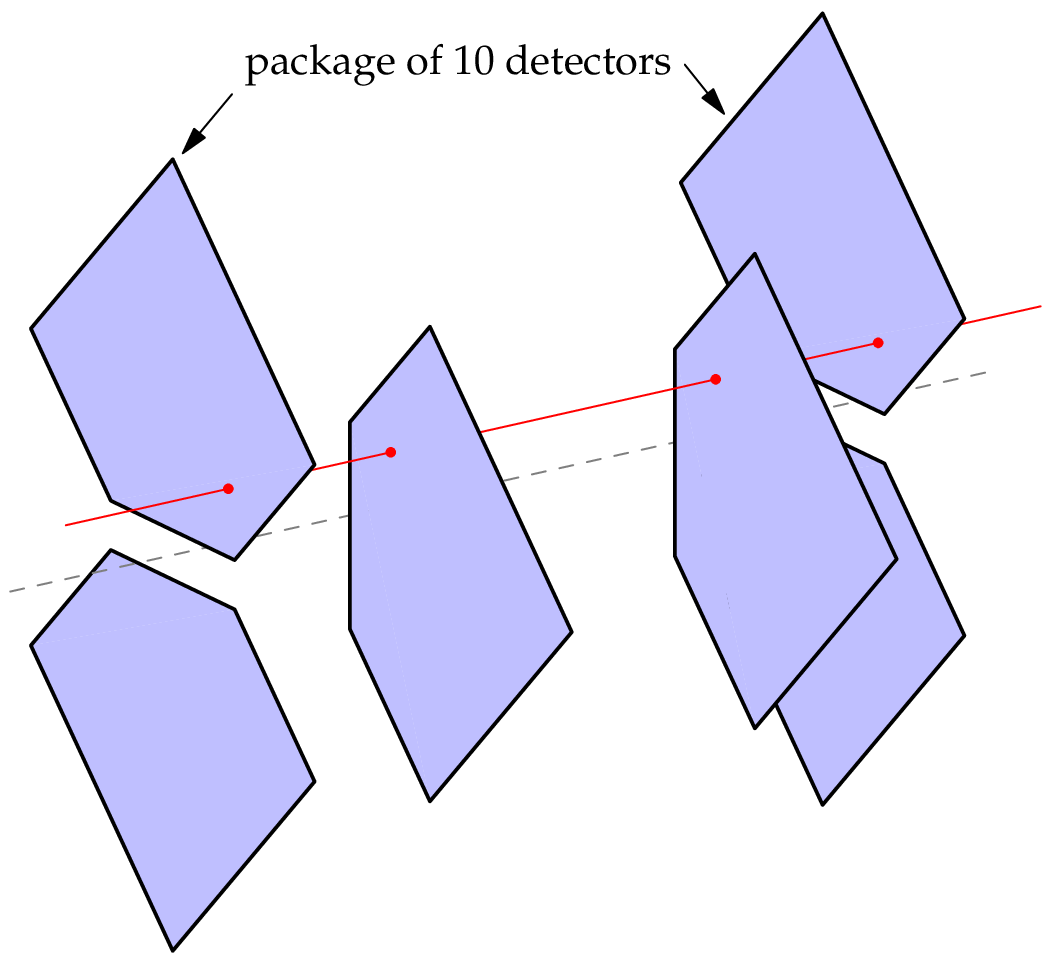}
    \caption{Schematic view of a RP station with a track traversing the overlap
zone between top and horizontal detectors.}
    \label{fig:stationscheme}
\end{wrapfigure}
\begin{figure}[h!]
   \vspace*{-3mm}
    \centering
    \hspace*{1mm}
    \includegraphics[width=83mm]{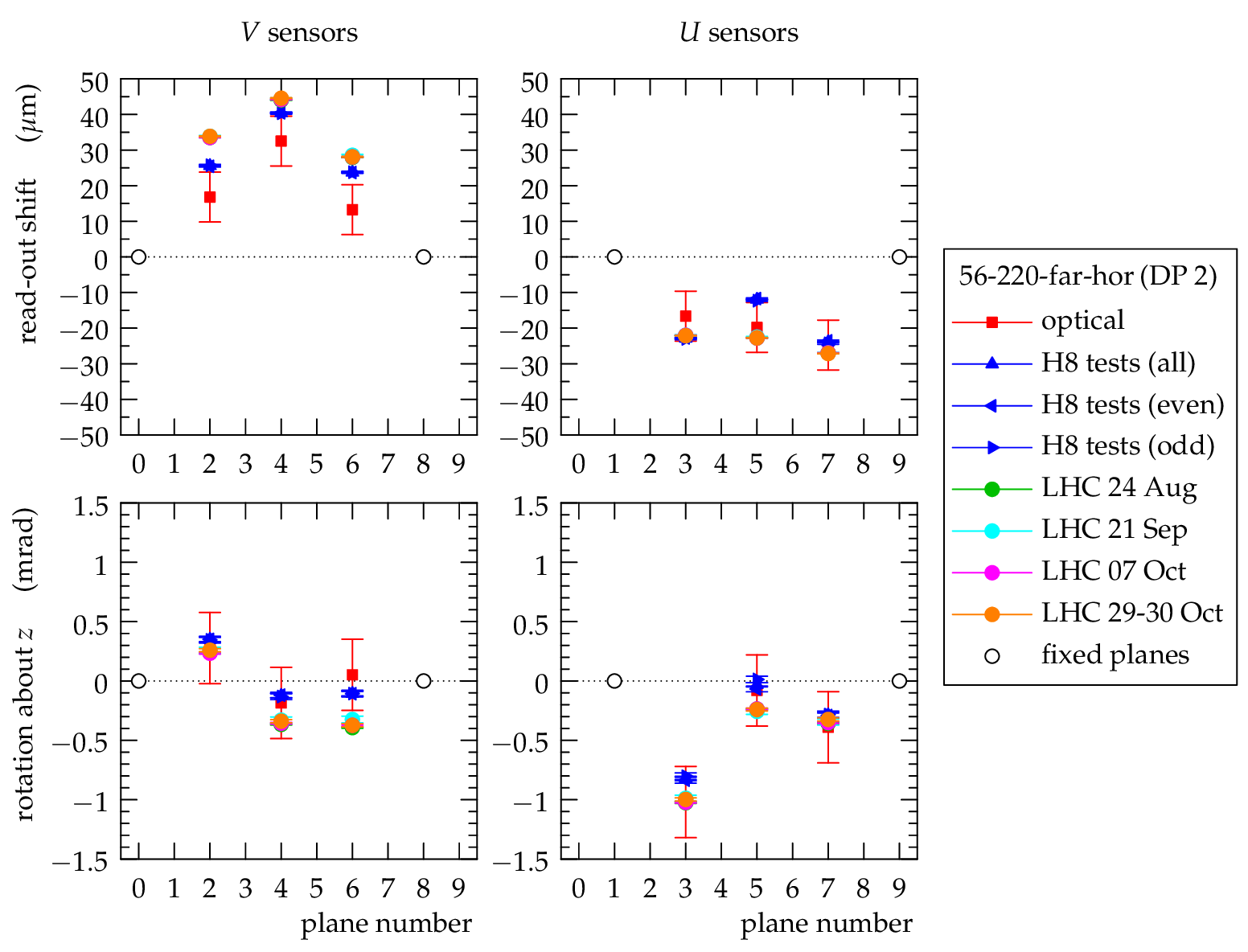}
    \vspace*{-5mm}
    \caption{Comparison of different alignment result for one example detector 
package. ``H8'' refers to a test beam alignment before installation, ``LHC'' 
to an alignment with LHC physics data in 2010, ``optical'' to the 
metrology measurement during assembly.}
    \label{fig:trackbasedresult}
\end{figure}

A detailed discussion of the software alignment methods is given in~\cite{jan}.
Most degrees of freedom of the relative alignment between the RP sensors can 
be obtained from the reconstructed tracks. The underlying idea is that sensor 
misalignments give raise to residuals, i.e. the distances of the measured hit 
positions from the fitted tracks. This technique is sensitive to shifts (along
the sensor read-out direction, i.e. perpendicular to the microstrips)
and rotations of individual detector planes relative to each other, but 
not to global shifts or rotations. The transverse overlap between vertical 
and horizontal detectors (Fig.~\ref{fig:stationscheme}) establishes the 
alignment between the mechanically independent 3 pots of a RP unit.
Fig.~\ref{fig:trackbasedresult} shows a comparison of the track-based alignment results for an example
RP with the optical metrology data.

\begin{figure}[h!]
   \vspace*{-3mm}
    \centering
    \hspace*{-2mm}
    \includegraphics[width=84mm]{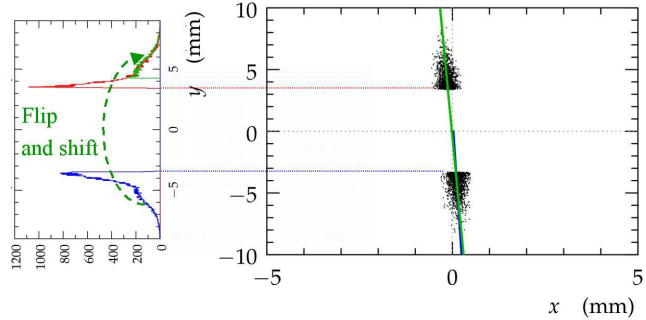}
    \vspace*{-8mm}
    \caption{Illustration of RP station alignment based on the symmetry of the
elastic scattering hit distribution. Right: distribution of track intercepts in
a scoring plane between the near and far units of a RP station. Note that the 
tilt of the vertical band is mainly caused by optics imperfections. Left: 
projection of (right) onto $y$. The symmetry line of the vertical distribution 
(indicating the position of the beam centre) is found by inverting the sign of 
the $y < 0$ part (excluding the region near
the acceptance cut) and shifting it until it coincides with the $y > 0$ part.}
    \label{fig:elasticprofile}
\end{figure}

The global misalignment modes (e.g. common shifts or rotations of entire 
stations w.r.t. the beam) 
are inaccessible to the track-based techniques but can be constrained 
by exploiting known symmetries of certain physics processes.
A prominent example is elastic scattering with its clean hit distribution
in the vertical detectors (Fig.~\ref{fig:elasticprofile}). The horizontal and
vertical symmetries of the distribution lead to an alignment precision of 
better than 10\,$\mu$m horizontally and about 20\,$\mu$m vertically.
%



\begin{thebibliography}{9}   
\bibitem{TOTEM} TOTEM Collaboration: Technical Design Report, 
CERN-LHCC-2004-002, addendum CERN-LHCC-2004-020. G. Anelli et al.: The TOTEM
Experiment at the CERN Large Hadron Collider, JINST 3 (2008) S08007.
\bibitem{usecases} M. Dutour et al.: TOTEM Roman Pots Control System Use Cases Specification, EDMS 937276r9.0, 2010.
\bibitem{cca} R. Assmann et al.: Application Software for the LHC Collimators 
and Movable Elements, EDMS 826861, 2007.
\bibitem{fesa} FESA: www.cern.ch/project-fesa .
\bibitem{middle}
S. Redaelli, A. Masi: Middle-Level Interface to Control Movable
Devices like LHC Collimators, EDMS 934341, 2008.
\bibitem{pxi} {\bf P}ci e{\bf X}tensions for {\bf I}nstrumentation, \verb| www.pxisa.org |
\bibitem{review} M. Deile et al.: Revision of the Movement Control for the TOTEM
and ALFA Roman Pots, EDMS 1141700v1.3, 2011.
\bibitem{lvdt} Macro Sensors: LVDT Basics, Techn. Bull. 0103, 2003.
\bibitem{interlock} M. Deile et al.: The TOTEM Interlock Logic in 2011: Specification and Test Results, EDMS 1141699, 2011.
\bibitem{collim} C. Bracco et al., ``Collimators and Beam Cleaning: 
First Results and Future Plans'', 
Proceedings from the LHC beam commissioning workshop, 
Evian 2010, \verb| https://espace.cern.ch/acc-tec-sector/Evian/ |\\
\verb| Proceedings.aspx |.
\bibitem{blm} R. Appleby et al.: Calibration of the BLM Response at 450\,GeV by
Beam Scraping with Roman Pots, CERN-TOTEM-Note-2010-001, and references therein.
\bibitem{jan} J. Ka\v{s}par: Elastic Scattering at the LHC, PhD thesis, Charles
University Prague, being submitted.
\end{thebibliography}
\end{document}